\documentclass[aps,floatfix,prd,twocolumn]{revtex4-2}
\usepackage{graphicx}
\usepackage{bm}
\usepackage{amsmath}
\usepackage{color}
\usepackage{tabularx}

\def\D{\bar D}

\def\u{\bar u}
\def\d{\bar d}
\def\s{\bar s}
\def\c{\bar c}
\def\b{\bar b}
\def\n{\bar n}

\def\dka{D^-K^+}
\def\dkb{\D^0K^0}
\def\dkc{\D^0K^+}
\def\dkd{\D^0K^0}
\def\dke{D^-K^+}
\def\dkf{D^-K^0}
\def\ddka{D^+D^-K^+}
\def\ddkb{D^+\D^0K^0}
\def\ddkc{D^0\D^0K^+}
\def\ddkd{D^0\D^0K^0}
\def\ddke{D^0D^-K^+}
\def\ddkf{D^+D^-K^0}
\def\bddka{\<D^+[\dka]\|O\|B^+\>}
\def\bddkb{\<D^+[\dkb]\|O\|B^+\>}
\def\bddkc{\<D^0[\dkc]\|O\|B^+\>}
\def\bddkd{\<D^0[\dkd]\|O\|B^0\>}
\def\bddke{\<D^0[\dke]\|O\|B^0\>}
\def\bddkf{\<D^+[\dkf]\|O\|B^0\>}
\def\bdxa{B^+\to D^+X, X\to \dka}
\def\bdxb{B^+\to D^+X, X\to \dkb}
\def\bdxc{B^+\to D^0X^+, X^+\to \dkc}
\def\bdxd{B^0\to D^0X, X\to \dkd}
\def\bdxe{B^0\to D^0X, X\to \dke}
\def\bdxf{B^0\to D^+X^-, X^-\to \dkf}
\def\tautau{\bm\tau_1\cdot\bm\tau_2}
\def\*{^{(*)}}
\def\brf{\mathcal B}

\def\>{\big>}
\def\<{\big<}
\def\|{\big\vert}
\def\rf#1{(\ref{#1})}

\def\ds{\displaystyle}

\def\etal{\textit{et~al.~}}

\begin{document}

\title{Discriminating among interpretations for the $X(2900)$ states}
\author{T.\,J.\,Burns}
\affiliation{Department of Physics, Swansea University, Singleton Park, Swansea, SA2 8PP, UK.}
\author{E.\,S.\,Swanson}
\affiliation{Department of Physics and Astronomy,
University of Pittsburgh,
Pittsburgh, PA 15260,
USA.}

\begin{abstract}
We make predictions for the production and decays of $X(2900)$ states, and their possible charged partners, in $B^+$ and $B^0$ decays, considering a number of competing models for the states, including triangle diagrams mediated by quark exchange or pion exchange, and resonance scenarios including molecules and tetraquarks.  Assuming only isospin symmetry and the dominance of colour-favoured weak decays, we find characteristic differences in the predictions of the different models. Future experimental studies can therefore discriminate among the competing interpretations for the states.
\end{abstract}

\maketitle

\section{Introduction}
The LHCb collaboration has recently reported a very prominent structure in the $D^-K^+$ spectrum in $B^+\to\ddka$ decays \cite{Aaij:2020hon,Aaij:2020ypa}. In an amplitude model in which the structure is described by Breit-Wigner resonances, their fit includes two states, $X_0(2900)$ with $J^P=0^+$,
\begin{align}
M &= 2.866 \pm 0.007 \pm 0.002 \textrm{ GeV}, \\\Gamma& = 57 \pm 12 \pm 4 \textrm{ MeV},
\end{align}
and $X_1(2900)$ with  $J^P=1^-$,
\begin{align}
M &= 2.904 \pm 0.005 \pm 0.001\textrm{ GeV},\\ \Gamma  &= 110 \pm 11 \pm 4 \textrm{ MeV}.
\end{align}
The $\dka$ decay implies the exotic flavour structure $ud\s\c$. If confirmed as genuine resonances, the states could be interpreted as exotic hadrons of either a molecular or tetraquark nature \cite{Molina:2010tx,Karliner:2020vsi,Hu:2020mxp,He:2020jna,Wang:2020xyc,Zhang:2020oze,Lu:2020qmp,Liu:2020nil,Chen:2020aos,He:2020btl,Huang:2020ptc,Xue:2020vtq,Molina:2020hde,Agaev:2019wkk,Agaev:2020nrc,Albuquerque:2020ugi,Mutuk:2020igv}. But as noted in the experimental analysis, and discussed elsewhere \cite{Liu:2020orv,Burns:2020epm,Chen:2020eyu}, a more prosaic explanation is also possible: the states could arise through triangle diagrams with $\D^*K^*\to D^-K^+$ or $\D_1 K\to \dka$ scattering.

Our own contribution to this discussion has been to demonstrate that the more prosaic possibility can give a good fit to experimental data \cite{Burns:2020epm}. The more exciting interpretation with exotic resonances, however, gives a marginally better fit to data, though we cannot really discriminate between the two.

The situation is very different compared to another candidate with exotic flavour, the $X(5568)$ observed at the D\O{} experiment \cite{D0:2016mwd}. We and others showed that both the triangle mechanism and more exotic molecular or tetraquark interpretations hopelessly fail in that case~\cite{Burns:2016gvy,Guo:2016nhb}. Subsequent experiments searched for, and did not find, $X(5568)$ \cite{Aaij:2016iev,Sirunyan:2017ofq,Aaltonen:2017voc,Aaboud:2018hgx}.

In this paper we do not advocate a particular model for the $X(2900)$ states, but instead give predictions for experiment which can discriminate among models. We also derive more general predictions which apply to all models.

We notice, for example, that as well as the discovery mode $\bdxa$, the $X(2900)$ states may also be seen in $\bdxb$. Experimental observation of the latter mode could discriminate among models, since its branching fraction compared to the discovery mode differs according to the nature of the
states. 

In addition to the above two modes, the $X(2900)$ states could also be seen in $\bdxe$ and $\bdxd$. Once again, predictions for these other modes depend  on the nature of the $X(2900)$ states, and so offer useful experimental tests.

Depending on their nature, the $X(2900)$ states may also be accompanied by charged partners in $\bdxc$ or $\bdxf$. The existence or otherwise
of these partners, and their branching fractions, can further discriminate among
models.

Our arguments are very simple and only rely on physical principles which are demonstrably satisfied by experimental data, namely that colour-favoured topologies dominate $B$ meson decay, and that strong interaction
vertices approximately satisfy isospin symmetry. 

\begin{figure*}
\includegraphics[width=0.8\textwidth]{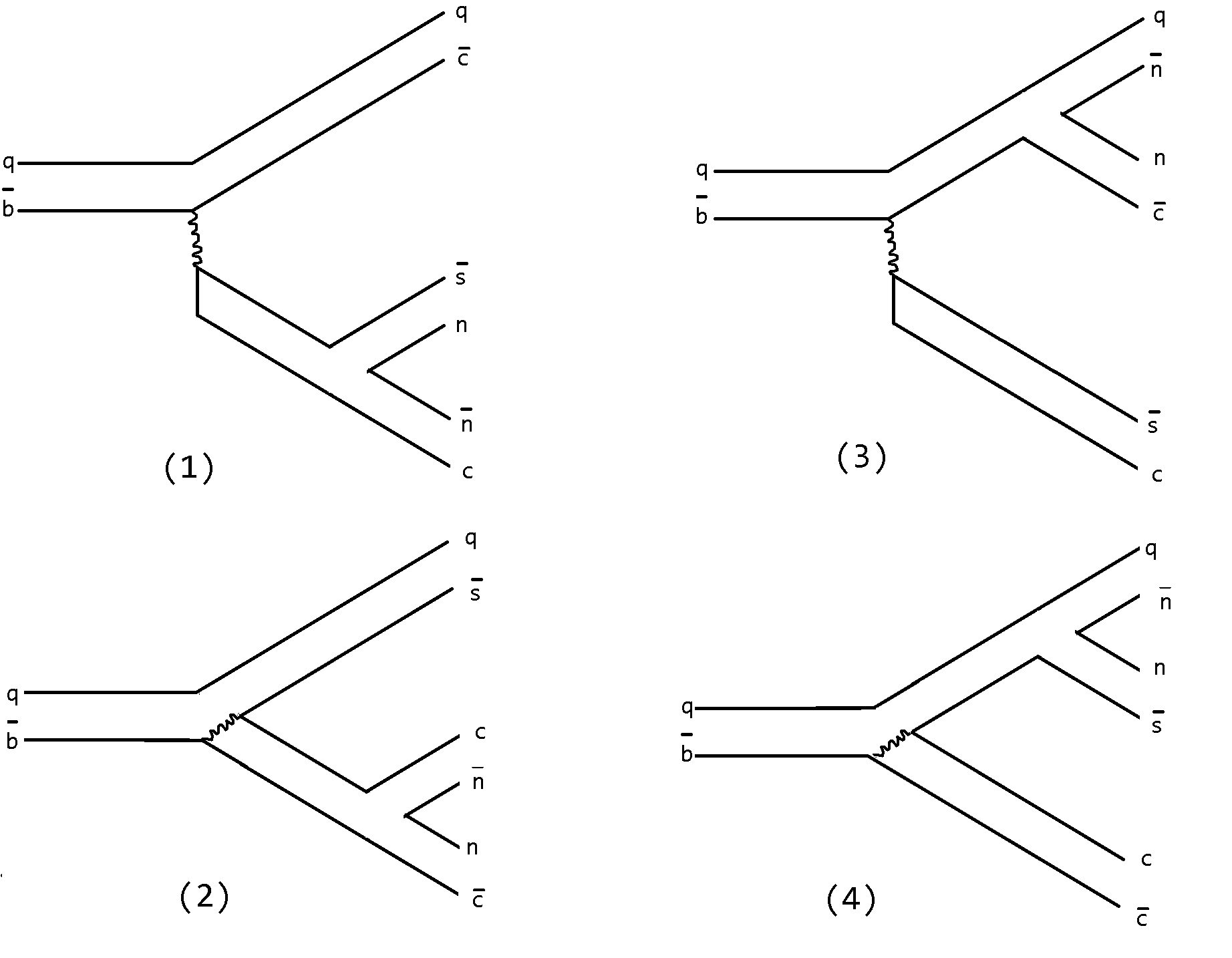}
\caption{Possible topologies for the decay of a $B=q\b$ state via the Cabibbo-favoured transition $\b\to \c(c\s)$ along with the creation of an isoscalar $n\n=(u\u+d\d)/\sqrt 2$ pair.}
\label{fig:top}
\end{figure*}
While this paper was in preparation, we noticed another paper which also discusses predictions for the $X(2900)$ states in $B$ decays  \cite{Chen:2020eyu}. Our approach and conclusions are, however, different. Chen~\etal derive relations which apply in the particular case that the $X(2900)$ states are resonances with fixed isospin, either $I=0$ or $I=1$. Here we additionally consider the case of mixed isospin, as well as several variations on the alternative scenario in which the states arise through triangle diagrams. We also derive a number of more general results that apply to all models.

Where the physics of our paper overlaps with theirs, we reach different conclusions. Whereas Chen~\etal quote  $B\to D X$ branching fractions, we point out there is no way to extract such branching fractions from the current experimental data. Instead we derive lower limits, concluding that the  $B\to D X$ branching fractions are larger by at least a factor of two compared to those of Chen~\etal

Furthermore, whereas the branching fractions for $B\to D X$  cannot be obtained directly, those of $B\to D X, X\to \D K$ can. We give formulae for these in our model, and quantify the predictions in terms of fit fractions. Some of these predictions are universal whereas others can discriminate among models.

In Section~\ref{sec:weak} we introduce the basic idea, identifying the dominant weak decay topology and its implications for the production of the $X(2900)$ states. In Sections~\ref{sec:general} and \ref{sec:discriminating} we derive relations among the matrix elements for $B\to \D X, X\to \D K$ transitions, firstly identifying general results, and then specialising to other results which apply separately to the different models. From these relations among matrix elements we derive, in Section~\ref{sec:bf}, predictions for experimental fit fractions, identifying in particular any modes that are particularly useful for discriminating among models. In Section~\ref{sec:tb} we derive lower limits on the $B\to D X$ branching fractions, valid only for the resonance interpretation. We conclude in Section~\ref{sec:conclusion}.

\section{Weak decay topologies}
\label{sec:weak}

\begin{figure*}
\includegraphics[width=\textwidth]{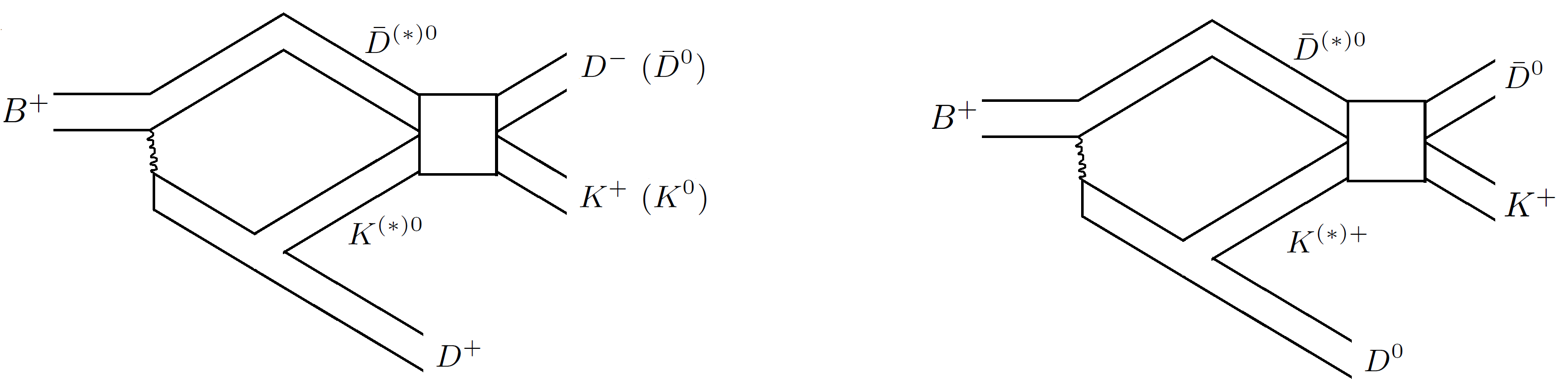}
\caption{Production of $X(2900)$ states (left) and their possible charged partner states (right) in $B^+$ decays. The labels ``$\bar D^{(*)0} K^{(*)0}$'' on the intermediate states refer to flavour only; the states could include any possible spin or orbital configurations such as $\D^{*0}K^{*0}$ or $D_1^0 K^0$.}
\label{fig:topb}\end{figure*}
In terms of quark 
flavours, the $B^+\to\ddka$ transition is \[u\b\to(c\d)(d\c)(u\s)\]
and so arises from a Cabibbo-favoured weak transition $\b\to \c(c\s)$ along with the creation of a $d\d$ pair from the
strong interaction. Using flavour considerations we may relate this transition to others where the created pair is $u\u$ rather than $d\d$, and where the initial state is $B^0$ rather than $B^+$.

We thus consider the flavour structure for decays of a
generic $B = q\b$ state (either $B^+=u\b$ or $B^0=d\b$), and
in which the pair created by the strong interaction is an
isoscalar mixture $n\n=(u\u+d\d)/\sqrt 2$. As a reminder, we summarise the quark flavour compositions of the hadrons involved:
\begin{align}
B^0&=d\b&B^+&=u\b\\
D^0&=c\u& D^+&=c\d\\
D^-&=d\c&\D^0&=u\c\\
K^0&=d\s&K^+&=u\s
\end{align}
The possible transitions are shown in Fig.~\ref{fig:top}. Topologies (1) and (3) involve ``external'' $W$ emission, whereas topologies (2)
and (4) have ``internal'' $W$ emission. Since they are observed in $B^+\to D^+ X$,  the $X(2900)$ states can only be produced via either of topologies (1) or (2), since only these have an outgoing $D^+$ meson (taking $n=d$). In each case the remaining two legs of the diagram combine to form the $X(2900)$ states. 

Ref.~\cite{Bondar:2020eoa} also identified the roles of what we call topologies (1) and (2), although their discussion has a very different focus from ours. They notice a charge asymmetry in $B\to D\D K$ decays and explain this as an effect of interference between topologies (1) and (2).

We notice, however, a critical difference between the two topologies: whereas (1) is colour-favoured, (2) is colour-suppressed. Empirically in two-body $B$ decays the branching fractions for colour-favoured transitions, such as $\D \* D_s\*$ and $\D\* D_{sJ}$, overwhelmingly dominate those of colour-suppressed transitions, such as $\eta_c K\*$, $J/\psi K\*$, and $ \chi_{cJ}K\*$. (Some examples are shown on the right axis of Fig.~\ref{fig:bf} -- note the logarithmic scale.) We therefore assume that the production of $X(2900)$ states is also dominated by the colour-favoured topology (1).

We further assume that the strong-pair creation vertex respects isospin, meaning that the $DK$ pair has isospin zero. With this assumption, we obtain the following flavour decompositions for the three-body states arising from $B^+$ and $B^0$ decays, respectively
\begin{align}
\|\phi^+\>&=\frac{1}{\sqrt 2}\left(\|\ddkb\>-\|\ddkc\>\right)\label{eq:phip}\\
\|\phi^0\>&=\frac{1}{\sqrt 2}\left(\|\ddkf\>-\|\ddke\>\right)\label{eq:phin}
\end{align}
Notice that the compositions are related by the replacement $\D^0\to D^-$.

Recalling that the $X(2900)$ states are observed in $B^+\to D^+X$, from \rf{eq:phip} we conclude that they are produced via the component with flavour $\dkb$. As an example we show in the left panel of Fig.~\ref{fig:topb} a diagram corresponding to $B^+\to D^+X, X\to D^-K^+$, via topology~(1). We emphasise that the intermediate state in the loop has flavour $\D^0 K^0$, not $D^- K^+$. However we make no assumption about the other quantum numbers of the intermediate state; in particular, it could correspond to $\D^{*0}K^{*0}$ or $D_1^0 K^0$, or some linear combination of any states with this flavour combination.

Other production and decay modes are also possible. In particular, for $B^+$ decays we identify the following processes:
\begin{align}
B^+&\to D^+X,~X\to \dka,\label{eq:a}\\
B^+&\to D^+X,~X\to \dkb,\label{eq:b}\\
B^+&\to D^0X^+,~X^+\to \dkc\label{eq:c}.
\end{align}
The first two arise from the first term in $\phi^+$. Here the $X(2900)$ states are produced via an intermediate state with flavour $\dkb$, and decay either into $\dka$ (as in their discovery mode) or $\dkb$. These are shown in the left panel of Fig.~\ref{fig:topb}. 

The third process comes from the second term in $\phi^+$. Here the possible charged partners of the $X(2900)$ states would be produced through an intermediate state with flavour $\dkc$, and would decay into $\dkc$. (See the right panel of Fig.~\ref{fig:topb}.)

The three analogous processes in $B^0$ decays are
\begin{align}
B^0&\to D^0 X,~X\to \dkd,\label{eq:d}\\
B^0&\to D^0 X,~X\to \dke,\label{eq:e}\\
B^0&\to \D^+ X^-,~X^-\to \dkf.\label{eq:f}
\end{align}
Referring to equation~\rf{eq:phin}, in this case the neutral $X(2900)$ states would be produced through an intermediate state with flavour $\dka$, and the charged states through $\dkf$.

Our key observation is that the matrix elements for all of these processes are related. We distinguish relations which are very general (Section \ref{sec:general}) from those which depend on the model for the $X(2900)$ states and so can be used to discriminate among them (Section \ref{sec:discriminating}).

\section{General results}
\label{sec:general}

As noted above, we will assume that the production of the $X(2900)$ states and their possible partners is driven by the colour-favoured topology, such that the flavour composition of the intermediate three-body state $D\D K$ state is given by equations \rf{eq:phip} and \rf{eq:phin} for $B^+$ and $B^0$ decays, respectively. The $\D K$ components in those wavefunctions combine, as in Fig~\ref{fig:topb}, to generate the $X(2900)$ states, which are observed in another (in general, different) $\D K$ combination. We emphasise that only the flavour of the intermediate $\D K$ state is specified; it could correspond, for example, to $\D^*K^*$, $\D_1 K$, or some linear combination of these or other states with the same flavour.

We introduce an operator $O$ to effect the transition from the intermediate ``$\D K$'' state to the final $\D K$ state (the rectangle in Fig.~\ref{fig:topb}). Clearly, the nature of the $X(2900)$ states is intimately related to the properties of the operator $O$. We later consider the different models and their corresponding operators. In this section we instead derive more general results which apply to all models and which require minimal assumptions for $O$. 

We denote the flavour part of the matrix element for a generic transition of the type 
 \begin{align}
B\to DX,~X\to \D K,
\end{align}
as 
\begin{align}
\<D[\D K]\|O\|B\>
\end{align}
For $B^+$ ($B^0$) decays, this is the matrix element of the operator $O$ between the intermediate state $\phi^+$ ($\phi^0$) and the final $D\D K$ state, noting that $O$ acts only the $\D K$ components of each. For the transitions in eqns \rf{eq:a}-\rf{eq:f} we have, respectively,
\begin{align}
\bddka&=\frac{1}{\sqrt 2}\<\dka\|O\|\dkb\>\label{eq:oa}\\
\bddkb&=\frac{1}{\sqrt 2}\<\dkb\|O\|\dkb\>\label{eq:ob}\\
\bddkc&=-\frac{1}{\sqrt 2}\<\dkc\|O\|\dkc\>\label{eq:oc}\\
\bddkd&=-\frac{1}{\sqrt 2}\<\dkd\|O\|\dka\>\label{eq:od}\\
\bddke&=-\frac{1}{\sqrt 2}\<\dke\|O\|\dka\>\label{eq:oe}\\
\bddkf&=\frac{1}{\sqrt 2}\<\dkf\|O\|\dkf\>\label{eq:of}
\end{align}
Note the following relation 
\begin{align}
\<D^+[\dka]\|O\|B^+\>&=-\<D^0[\dkd]\|O\|B^0\>\label{eq:gen}
\end{align}
from which, using experimental data from the $X(2900)$ discovery mode (left-hand side), we will later predict the fit fraction for a new production and decay mode (right-hand side). 

We now impose the further assumption, which is still very general and which applies to almost all the models we consider later, that the operator  $O$ is a scalar with respect to isospin symmetry. For $\D K$ states coupled to total isospin $I$ and third component $I_3$, the operator satisfies
\begin{align}
O\|(\D K)_I^{I_3}\>=\lambda_I\|(\D K)_I^{I_3}\>\label{eq:we}
\end{align}
Namely, it conserves $I$ and $I_3$ and, from the Wigner-Eckart theorem, its eigenvalues $\lambda_I$ depend on $I$, but not~$I_3$. In this way we can express the flavour dependence of all transitions  in terms of two parameters, $\lambda_0$ and $\lambda_1$. 

The states in this isospin basis are
\begin{align}
\|(\D K)^0_0\>&=\frac{1}{\sqrt 2}\left(\|\D^0K^0\>-\|D^-K^+\>\right)\\
\|(\D  K)_1^+\>&=\|\D^0 K^+\>\\
\|(\D K)_1^0\>&=\frac{1}{\sqrt 2}\left(\|\D^0K^0\>+\|D^-K^+\>\right)\\
\|(\D K)_1^-\>&=\|D^- K^0\>
\end{align}
where the superscript labels the electric charge, which is equivalent to $I_3$.

From the preceding equations we find
\begin{align}
\bddka&=-\bddkd=\frac{\lambda_1-\lambda_0}{2\sqrt 2}\label{eq:wea}\\
\bddkb&=-\bddke=\frac{\lambda_1+\lambda_0}{2\sqrt 2}\label{eq:web}\\
\bddkc&=-\bddkf=-\frac{\lambda_1}{\sqrt 2}\label{eq:wec}
\end{align}
The first of these recovers the previous relation. The other two are additional relations among new production and decay modes for the $X(2900)$ states and their possible charged partners.  We translate these relations into fit fractions in Section~\ref{sec:bf}.

\section{Discriminating among models}
\label{sec:discriminating}

At this point the six production and decay modes of the $X(2900)$ states and their charged partners are expressed in terms of the eigenvalues $\lambda_1$ and $\lambda_0$ of the operator~$O$. We now consider various models for the $X(2900)$ states, which effectively fixes $O$ and its eigenvalues, and gives further relations among the six production and decay modes. These relations are specific to each model, and so give experimental tests which can discriminate among possible interpretations of the $X(2900)$ states.

The results (which we explain further below) are summarised in Table~\ref{tab:me}. In the first row we give the general expressions for the matrix elements \rf{eq:wea}-\rf{eq:wec}. The remaining entries in the table, which give the corresponding matrix elements for particular models, can be obtained from these general expressions by specifying $\lambda_1$ and $\lambda_0$.

\begin{table*}

\begin{tabularx}{\textwidth}{XlXXl}
\hline\noalign{\medskip}
Model& Operator	\phantom{-----}&$\bddka$&$\bddkb$&$\bddkc$\\
	&&$=-\bddkd$&$=-\bddke$&$=-\bddkf$\\
\noalign{\medskip}\hline\noalign{\medskip}
General&$O$&$\ds\frac{\lambda_1-\lambda_0}{2\sqrt 2}$&$\ds\frac{\lambda_1+\lambda_0}{2\sqrt 2}$&$\ds-\frac{\lambda_1}{\sqrt 2}$\\
\noalign{\medskip}
Triangle, QE &$Q$&$\ds\frac{1}{\sqrt 2}$ & 0 & $\ds-\frac 1{\sqrt 2}$\\
\noalign{\medskip}
Triangle, OPE&$\tautau$&$\ds\sqrt 2$&$\ds-\frac 1{\sqrt 2}$&$\ds-\frac 1{\sqrt 2}$\\
\noalign{\medskip}
Triangle, EFT&$T$&$\ds\sqrt 2C_1$&$\ds\frac{C_0-C_1}{\sqrt 2}$&$\ds-\frac{C_0+C_1}{\sqrt 2}$\\
\noalign{\medskip}
Resonance, $I=0$&$P_0$&$\ds-\frac 1{2\sqrt 2}$&$\ds\frac 1{2\sqrt 2}$&0\\
\noalign{\medskip}
Resonance, $I=1$&$P_1$&$\ds\frac 1{2\sqrt 2}$&$\ds\frac 1{2\sqrt 2}$&$\ds-\frac 1{\sqrt 2}$\\
\noalign{\medskip}
Resonance, $I$ mixed&$P_\theta$&$\ds\frac{1}{2\sqrt 2}(\sin^2\theta-\cos^2\theta)$&$\ds\frac{1}{2\sqrt 2}(\sin\theta\pm\cos\theta)^2$ (*)&\\
\noalign{\medskip}
\hline\noalign{\smallskip}
\end{tabularx}
\caption{Matrix elements for $B\to DX, X\to \D K$ transitions in various models. For an explanation of the entry marked $(^*)$, see equations~\rf{eq:va} and \rf{eq:vb} and the subsequent text.}
\label{tab:me}
\end{table*}

We consider two classes of model. Firstly, the $X(2900)$ states could arise due to triangle diagrams, through scattering processes such as $\D^*K^*\to \dka$ or $\D_1 K\to \dka$. {We consider a number of possible models for the scattering process:} quark exchange (Sec.~\ref{sec:tra}), pion exchange (Sec.~\ref{sec:trb}), and effective field theory interactions (Sec.~\ref{sec:trc}). In these cases, the operator $O$ represents a direct scattering process from an intermediate flavour state (such as $\dkb$) to a final state (such as $\dka$). {We are assuming in these scenarios that the interactions are sufficiently weak as to be described by the perturbative, one-loop triangle diagram.}

In the second class of models, the $X(2900)$ states are resonances with $I=0$, $I=1$ {or mixed isospin} (Sec.~\ref{sec:trd}). Our results do not distinguish the dynamical origin of these resonances which could be, for example, molecular or tetraquark in nature. Unlike the triangle diagram scenario, {here the interactions are strong and non-perturbative.} In this set up the operator $O$ does not represent a direct scattering, but instead projects out the {isospin} components of the intermediate state (such as $\dkb$) and final state (such as $\dka$). 

{Note that the same interactions we consider for the triangle scenario -- quark exchange, pion exchange, and effective field theory interactions -- could, if sufficiently strong, give rise to resonances. We are making no claim as to which scenario (triangle versus resonance) is most relevant to each type of interaction. The reason we consider different interactions separately within the  triangle scenario is just that, in that case, they each have a different pattern of production and decay branching fractions. In the resonance scenario, on the other hand, the pattern of branching fractions is sensitive to the isospin of the state, but not to the underlying interactions.}

\subsection{Triangle with quark exchange}
\label{sec:tra}

The scattering of mesons can be understood as arising from the pair-wise interactions of their quark constituents~\cite{Barnes:1991em,Swanson:1992ec,Barnes:1992ca,Barnes:2000hu}. In this approach the colour structure of the interaction potential demands quark exchange (QE) between the mesons. Hence the flavour dependence is given by an operator $Q$ satisfying
\begin{align}
Q\|\dka\>&=\|\dkb\>\\
Q\|\dkb\>&=\|\dka\>\\
Q\|\dkc\>&=\|\dkc\>\\
Q\|\dkf\>&=\|\dkf\>
\end{align}

The matrix elements in this model  are shown in Table~\ref{tab:me}. These can be obtained directly from above, or by using the general expressions, noting that the relevant eigenvalues are $\lambda_1=1$ and $\lambda_0=-1$.

We note two interesting features of this model, which can be tested in experiment. Firstly, in $B^+\to D^+X$ decays the $X(2900)$ states should appear in $X\to \dka$ (consistent with experiment) but not $X\to\dkb$. The pattern is opposite for $B^0\to D^0X$, for which the $X(2900)$ states would appear in $X\to \dkb$ but not $X\to\dka$. Recalling equations \rf{eq:phip} and \rf{eq:phin}, notice that the final states which arise from the triangle mechanism are not produced directly via the colour-favoured mechanism; on the other hand, states which are produced directly via the colour-favoured mechanism, are forbidden through the triangle mechanism.


A second interesting feature  is that the triangle diagrams would also produce charged analogues of the $X(2900)$ signals. Comparing the first and last columns, the modes $\bdxc$ and $\bdxf$ would have the same rate as the $X(2900)$ discovery mode. (We translate this into fit fractions in Section~\ref{sec:bf}.)

{As noted above, we are assuming here that the interactions are sufficiently weak as to be described by the perturbative, one-loop triangle diagram. If quark-exchange interactions are strong enough to generate a resonance, the  results of Section~\ref{sec:trd} apply.}

\subsection{Triangle with one-pion exchange}
\label{sec:trb}

Scattering processes $\D^*K^*\to \D K$ or $\D_1 K\to \D K$
can also arise from the exchange of light mesons. For $\D^*K^*\to \D K$ the amplitude is presumably dominated by
one-pion exchange (OPE). (There is no pion-exchange diagram
for $\D_1 K\to \D K$.)

In this case the flavour dependence is given by the operator $\bm\tau_1\cdot\bm\tau_2$, where $\bm\tau_1$ and $\bm\tau_2$ act on the mesons with flavour $\D$ and $K$, respectively. Its eigenvalues are $\lambda_1=1$, $\lambda_0=-3$, from which we obtain the matrix elements in the third row of the table. {The results apply in the limit that OPE is essentially perturbative; molecular states bound by OPE would instead be described by the results of Section~\ref{sec:trd}.}

In this scenario all six production and decay modes are available but, comparing the three columns, the $B^+$ discovery mode and its $ B^0$ analogue are larger by a factor of 2 in amplitude (or 4 in rate) compared to the other modes. (We give fit fractions in Section~\ref{sec:bf}.)

Notably, in this model the neutral $X(2900)$ states would be seen in modes  (see the middle column) which are forbidden by the QE mechanism discussed previously, specifically $\bdxb$ and $\bdxe$. The model also predicts charged analogues of the $X(2900)$ states (last column).

As well as pions, the exchange of other light mesons is
also possible. In this case the 
flavour structure is more
complicated; we comment further on this possibility
in the next subsection.

\subsection{Triangle with EFT}
\label{sec:trc}

In the effective field theory (EFT) approach, the $\D^*K^*\to \D K$ scattering amplitude has a long-range contribution from pion-exchange, and a short-range contribution which is parametrised by contact terms which, in principle, are fit to data.
(Again, for $\D_1 K\to \D K$ there is no pion-exchange contribution.)
In general, the contact terms can include all operators which respect the appropriate symmetries, such as
the conservation of heavy-quark spin, and isospin. Here
we ignore the spin-dependence since we are interested in relations among scattering processes involving
hadrons with the same spins, but different
flavours.

The operators respecting the conservation of isospin
are the unit operator, and the $\bm\tau_1\cdot\bm\tau_2$ operator discussed
in the previous section. The resulting transition operator
therefore has the form
\begin{align}T=C_0+C_1\tautau,
\end{align}
where we can think of $C_0$ and $C_1$ as having absorbed
the spin-dependence for the particular scattering process
under comparison. (Note that $C_1$ would include contributions both from the contact term and long-range pion-exchange.)

The eigenvalues in this model are $\lambda_1=C_0+C_1$ and $\lambda_0=C_0-3C_1$, from which we obtain the matrix elements in Table~\ref{tab:me}. {Again, the results apply in the limit that the interactions are perturbative.}

Note that the results obtained in this way are equivalent, algebraically, to the more general results derived in the previous section. In both cases the matrix elements are parametrised in terms of two parameters -- either $\lambda_1$ and $\lambda_0$, or $C_1$ and $C_0$ -- and the underlying algebra of the matrix elements is identical. In this sense the EFT approach is less predictive than the others.

Nevertheless the chosen parametrisation in terms of $C_1$ and $C_0$ may still be useful. The usual EFT philosophy is that the short-distance physics cannot be derived from the underlying theory, and is instead parametrised by means of contact terms which are fit to data. An alternative approach is to model the short-distance physics, as with the long-range  physics, in terms of meson exchange. In this set-up, the terms $C_1$ and $C_0$ would, respectively, be associated with the exchange of isovector mesons ($\pi$, $\rho$, etc.) and isoscalar mesons ($\sigma$, $\eta$, $\omega$, etc.)

Finally we note that the $Q$ operator corresponding to QE scenario discussed previously is equivalent to the operator $T$ with $C_0 = C_1 = 1/2$.

\subsection{Resonance (molecule or tetraquark)}
\label{sec:trd}

We now move on to the very different scenario in which the $X(2900)$ states are resonances. We make no assumption about their underlying dynamics: for example they could be molecular in nature, or compact objects with constituent quark or diquark degrees of freedom. As such the results in this section apply equally to all such models.

To give meaning to the operator $O$ we consider the example shown in the left panel of Fig.~\ref{fig:topb}, where it mediates a transition from an intermediate state with flavour $\dkb$ to the final state with flavour $\dka$. For a resonance with flavour wavefunction $X$,  the matrix element factorises
\begin{align}
\<\dka\|O\|\dkb\>=\<\dka\|X\>\<X\|\dkb\>,\label{eq:fac}
\end{align}
so that $O$ is a projection operator
\begin{align}
O=\|X\>\<X\|.
\end{align}
This is very different to the previous examples where the $X(2900)$ states arise due to triangle diagrams. In particular, the factorisation of the matrix element into a product of two matrix elements is ultimately why, in the resonance scenario, there is an equivalent factorisation of the branching fractions,
\begin{align}
\brf(B\to DX, X\to \D K)=\brf(B\to DX)\brf(X\to \D K),\label{eq:factorise}
\end{align}
whereas the same does not happen in the triangle scenario. We return to this point in Section~\ref{sec:tb}.

If the $X(2900)$ resonances arise from interactions which respect isospin symmetry, they will have either $I=0$ or $I=1$. The corresponding projection operators are
\begin{align}
P_0&=\|(\D K)_0\>\<(\D K)_0\|,\\
P_1&=1-P_0=\sum_{I_3}\|(\D K)_1^{I_3}\>\<(\D K)_1^{I_3}\|.
\end{align}
The eigenvalues of $P_0$ are $\lambda_1=0$, $\lambda_0=1$, while those of $P_1$ are $\lambda_1=1$, $\lambda_0=0$. From these we obtain the matrix elements in Table~\ref{tab:me}.

Comparing the first two columns, we note that in both cases all four production and decay modes of the neutral $X(2900)$ states are possible, and have the same rate, regardless of their isospin. Hence the observation of modes other than the discovery mode cannot discriminate between $I=0$ and $I=1$, although it could discriminate between these and the alternative scenario of a triangle diagram with QE or OPE. 

The difference between the $I=1$ and $I=0$ hypotheses is that only in the former case would the neutral
$X(2900)$ states be  accompanied by charged partners. From the last column, these modes are enhanced by a factor of 2 in amplitude (4 in rate) compared to
the discovery mode, raising the realistic prospect of their observation in experiment. (We predict fit fractions in Section~\ref{sec:bf}.)

A third possibility (for the neutral $X(2900)$ states) is that isospin is not a good quantum number, and the wavefunctions are admixtures of
$I=1$ and $I=0$. This is most likely to be relevant
for the heavier $X_1(2900)$ in the molecular scenario, where unequal admixtures of $D^{*-}K^{*+}$ and $\D^{*0}K^{*0}$ in the wavefunction would arise from mass splittings between the corresponding thresholds. Such isospin splitting would be significant if the mass splittings between the thresholds is significant on the scale of the binding energy. The mechanism is analogous to the case of
$X(3872)$ \cite{Close:2003sg,Tornqvist:2004qy,Voloshin:2007hh}, and has also been discussed for $P_c(4457)$ \cite{Burns:2015dwa,Guo:2019fdo,Burns:2019iih}.

With current experimental uncertainties (on both $X(2900)$ and $K^{*+}$ masses) it is not possible to quantify the scale of any mixing. So instead we consider an arbitrary mixing angle:
\begin{align}
\|\theta\>=\cos\theta\|(\D K)_0^0\>+\sin\theta\|(\D K)_1^0\>.
\end{align}
The corresponding projection operator
\begin{align}
P_\theta=\|\theta\>\<\theta\|
\end{align}
does not satisfy equation \rf{eq:we}, as it couples $I=0$ and $I=1$ states. Consequently, the general expressions (in the top row of Table~\ref{tab:me})  which we used for all other cases cannot be used in this case. Nevertheless the required matrix elements can be obtained straightforwardly, for example with equations \rf{eq:oa}-\rf{eq:of} as a starting point. The results are
\begin{align}
\bddka&=\frac{1}{2\sqrt 2}(\sin^2\theta-\cos^2\theta)\\
\bddkd&=-\frac{1}{2\sqrt 2}(\sin^2\theta-\cos^2\theta)\\
\bddkb&=\frac{1}{2\sqrt 2}(\sin\theta+\cos\theta)^2\label{eq:va}\\
\bddke&=-\frac{1}{2\sqrt 2}(\sin\theta-\cos\theta)^2\label{eq:vb}
\end{align}
Note that whereas the first two relations satisfy equation \rf{eq:gen}, the second two do not satisfy the analogous relation \rf{eq:web}. This is because whereas equation \rf{eq:gen} is very general, equation \rf{eq:web} relies on the assumption that the operator conserves $I$, which it is not true of $P_\theta$. The violation of equation \rf{eq:web} would be an experimental indication of the mixed isospin nature of the $X(2900)$ states, as this does not happen in any other scenario. 

The matrix elements for the mixed isospin case are shown in the last row of  Table \ref{tab:me}. The entries for \rf{eq:va} and \rf{eq:vb} are indicated with $(^*)$ as a reminder that (unlike all other cases) they do not satisfy the corresponding relation in the top row of the table. 

A neutral state with mixed isospin may or may not have charged partners, depending on whether, in the absence of isospin breaking mass splittings, it would be an isosinglet, or the neutral member of an isotriplet. For this reason we make no entry in the last column.

So far we have considered matrix elements for the full transitions $B\to D X, X\to \D K$. For the resonance scenario, we can in addition consider the separate matrix elements for production $B\to DX$ and decay $X\to \D K$. This possibility follows from the factorisation \rf{eq:fac}, and is a consequence of the projective nature of the operator. The same does not apply to the triangle scenario, for which it is meaningless to separate the production and decay.

The production matrix elements ($B\to DX$) are easily obtained from the results in the table by taking the appropriate isospin-weighted combinations. For the $I=0$ interpretation we have
\begin{align}
\<D^+X\|O\|B^+\>=\<D^0 X\|O\|B^0\>=\frac{1}{2},\label{eq:nc}
\end{align}
whereas for $I=1$,
\begin{align}
&\<D^0X^+\|O\|B^+\>=-\sqrt 2\<D^+ X\|O\|B^+\>\nonumber\\
&=-\<D^+X^-\|O\|B^0\>=\sqrt 2\<D^0 X\|O\|B^0\>=-\frac{1}{\sqrt 2}.\label{eq:ncb}
\end{align}
These are consistent with the relations of Ref.~\cite{Chen:2020eyu}. 

For the isospin-mixed case, from the last two equations,
\begin{align}
\<D^+X\|O\|B^+\>&=\frac{\cos\theta +\sin\theta}{ 2},\\
\<D^0X\|O\|B^0\>&=\frac{\cos\theta -\sin\theta}{ 2},
\end{align}
so the relation between the two is
\begin{align}
\<D^+X\|O\|B^+\>=\tan\left(\theta+\frac{\pi}{4}\right)\<D^0X\|O\|B^0\>.\label{eq:ncc}
\end{align}

For the decay matrix elements ($X\to \D K$) the flavour dependence  is given by Clebsch-Gordan coefficients. For $I=0$,
\begin{align}
\<\dkb\|X\>=-\<\dka\|X\>=\frac{1}{\sqrt 2},\label{eq:ra}
\end{align}
and for $I=1$,
\begin{align}
&\<\dkc\|X^+\>=\<\dkf\|X^-\>\nonumber\\&=\sqrt 2\<\dkb\|X\>=\sqrt 2\<\dka\|X\>=1.\label{eq:rb}
\end{align}
For the mixed state,
\begin{align}
\<\dkb\|X\>&=\frac{\cos\theta +\sin\theta}{\sqrt 2},\\
\<\dka\|X\>&=\frac{-\cos\theta +\sin\theta}{\sqrt 2},
\end{align}
so the relation is
\begin{align}
\<\dkb\|X\>=-\tan\left(\theta+\frac{\pi}{4}\right)\<\dka\|X\>\label{eq:rc}.\end{align}

\begin{table*}

\begin{tabularx}{\textwidth}{XXXl}
\hline\noalign{\medskip}
& $\bdxa$&$\bdxb$&$\bdxc$\\
	&$\bdxd$&$\bdxe$&$\bdxf$\\
\noalign{\medskip}\hline\noalign{\medskip}
Triangle, QE&1 & 0 & 1\\
\noalign{\medskip}
Triangle, OPE&1&$\ds\frac 1{4}$&$\ds\frac 1{4}$\\
\noalign{\medskip}
Triangle, EFT&1&$\ds\frac{1}{4}\left(1-\frac{C_0}{C_1}\right)^2$&$\ds\frac{1}{4}\left(1+\frac{C_0}{C_1}\right)^2$\\
\noalign{\medskip}
Resonance, $I=0$&1&1&0\\
\noalign{\medskip}
Resonance, $I=1$&1&1&4\\
\noalign{\medskip}
Resonance, $I$ mixed
&1&$\ds\tan^2\left(\theta\pm\frac{\pi}{4}\right)$ (*)&\\
\noalign{\medskip}
\hline\noalign{\smallskip}
\end{tabularx}
\caption{The ratio $R$, defined in equation \rf{eq:r}, for $B\to DX, X\to \D K$ transitions in various models. For the entry marked~$(^*)$, the upper and lower signs are for the $B^+$ and $B^0$ decays, respectively.}
\label{tab:r}
\end{table*}

\section{Fit fractions}
\label{sec:bf}

We now give some quantitative predictions for experiment. The measured fit fractions
\begin{multline}
f(B^+\to D^+X, X\to D^-K^+)\\
=\frac{\brf(B^+\to D^+X, X\to D^-K^+)}{\brf(B^+\to\ddka)}
\end{multline}
are
\begin{align}
f=\left\{\begin{array}{ll}(5.6\pm 1.4\pm 0.5)\%,& X_0(2900)\\
(30.6\pm 2.4\pm 2.1)\%,& X_1(2900)\end{array}\right.
\end{align}
With the experimental branching fraction $\brf(B^+\to\ddka)$ \cite{Zyla:2020zbs} we get the branching fractions in the numerator,
\begin{multline}
\brf(B^+\to D^+X, X\to D^-K^+)\\
=\left\{\begin{array}{ll}(1.23\pm 0.42\pm 0.30)\times 10^{-5},& X_0(2900),\\
(6.73\pm 1.62\pm 1.60)\times 10^{-5},& X_1(2900).\end{array}\right.\label{eq:bft}
\end{multline}
(Note that the same numbers appear in Ref.~\cite{Chen:2020eyu}, although they quote these as $\brf(B^+\to D^+X)$ rather than $\brf(B^+\to D^+X, X\to D^-K^+)$. They similarly make predictions for other $\brf(B \to DX)$ branching fractions. In Section~\ref{sec:tb} we show that $\brf(B \to DX)$ fractions are larger by at least a factor of two compared to those quoted in Ref.~\cite{Chen:2020eyu}.)

We can now make predictions for branching fractions and fit fractions for other modes $B\to DX, X\to \D K$ in terms of the corresponding experimental numbers for $B^+\to D^+X, X\to D^-K^+$. A convenient quantity in this respect is the ratio of squared matrix elements
\begin{align}
R=\left(\frac{\<D[\D K]\|O\|B\>}{\bddka}\right)^2,\label{eq:r}
\end{align}
which follows immediately from Table~\ref{tab:me}, and which we summarise in Table~\ref{tab:r}.  The pattern of numbers $R$ is ultimately what discriminates among the predictions of different models.

\begin{table*}
\begin{tabularx}{\textwidth}{lXXXXXl}
\noalign{\medskip}\hline\noalign{\medskip}
& 
$B^+\to D^+X,$&$B^0\to D^0X,$&$B^+\to D^+X,$&$B^0\to D^0X,$&$B^+\to D^0X^+,$&$B^0\to D^+X^-,$\\
&$~~X\to\dka$&$~~X\to\dkd$&$~~X\to\dkb$&$~~X\to\dke$&$~~X^+\to\dkc$&$~~X^-\to\dkf$\\
\noalign{\medskip}\hline\noalign{\medskip}
$\brf(B\to D\D K)$&2.2$~\pm 0.7$&2.7$~\pm 1.1$&15.5$~\pm2.1$&10.7$~\pm 1.1$&14.5$~\pm 3.3$&7.5$~\pm 1.7$\\
\noalign{\medskip}\hline\noalign{\medskip}
$f(B\to DX, X\to \D K)$&&&&&&\\
\noalign{\medskip}
\quad Triangle, QE &30.6&23.2&0&0&4.6&8.3\\
\noalign{\medskip}
\quad Triangle, OPE&30.6&23.2&1.1&1.5&1.2&2.1\\
\noalign{\medskip}
\quad Triangle, EFT&30.6&23.2&$ 1.1\left(1-\frac{C_0}{C_1}\right)^2$&$ 1.5\left(1-\frac{C_0}{C_1}\right)^2$&$ 1.2\left(1+\frac{C_0}{C_1}\right)^2$&$ 2.1\left(1+\frac{C_0}{C_1}\right)^2$\\
\noalign{\medskip}
\quad Resonance, $I=0$&30.6&23.2&4.3&5.8&0&0\\
\noalign{\medskip}
\quad Resonance, $I=1$&30.6&23.2&4.3&5.8&18.6&33.4\\
\noalign{\medskip}
\quad Resonance, $I$ mixed&30.6&23.2&$ 4.3\tan^2\left(\theta+\frac{\pi}{4}\right)$&$ 5.8\tan^2\left(\theta-\frac{\pi}{4}\right)$&&\\
\noalign{\medskip}\hline\noalign{\medskip}
$\Delta f/f$ &0.1 &0.53&0.36&0.35&0.41&0.41\\
\noalign{\medskip}\hline\noalign{\medskip}
\end{tabularx}
\caption{The first row shows the experimental three-body branching fraction $\brf (B\to D\D K)$, in units of $10^{-4}$, from Ref.~\cite{Zyla:2020zbs}. The rest of the table shows fit fractions $f(B\to DX, X\to \D K)$ in percent. The first column is from experiment, and the remaining columns are predictions, obtained using equation \rf{eq:ff}. The fractional uncertainty $\Delta f/f$ is shown in the last row.}
\label{tab:fit}
\end{table*}

Ignoring small differences due to phase space, the relations between the predicted modes and the discovery mode are, for the branching fractions,
\begin{align}
\frac{\brf (B\to DX, X\to \D K)}{\brf(\bdxa)}=R\frac{\tau(B)}{\tau(B^+)},
\end{align}
and for the fit fractions,
\begin{multline}
\frac{f(B\to DX, X\to \D K)}{f(\bdxa)}\\=R\frac{\tau(B)}{\tau(B^+)}\frac{\brf(B^+\to \ddka)}{\brf(B\to D \D K)},\label{eq:ff}
\end{multline}
where here $\tau(B)$ is the lifetime of the $B^+$ or $B^0$ meson under consideration.

From the preceding two equations, and the $R$ values in Table~\ref{tab:r}, we may predict branching fractions and fit fractions for all  production and decay channels for the $X(2900)$ states and their possible partners, in all of the models considered. In order not to be overwhelmed with numbers, we give explicit predictions only for the heavier $X_1(2900)$ state; the corresponding numbers for $X_0(2900)$ are smaller by a factor 5.6/30.6. Similarly, we do not quote numbers for the branching fractions, but instead quote only fit fractions, since these are the quantities which are directly measured in experimental amplitude analyses, and which indicate (more directly than the branching fraction) how prominent a particular channel will feature in the amplitude fit.

The results are shown in Table~\ref{tab:fit}. 
In the first row we show the experimental branching fractions  $\brf (B\to D\D K)$, in units of $10^{-4}$, taken from Ref.~\cite{Zyla:2020zbs}. The rest of the table shows the fit fractions $f(B\to DX, X\to \D K)$, in percent. The numbers in the first column are those measured in experiment; the rest are predictions, obtained from equation \rf{eq:ff}. Within a given column, all that distinguishes one entry from another is the number $R$, which is shown in Table~\ref{tab:r}. Fractional uncertainties on $f$ are shown in the last row: the first is from experiment, the rest are obtained by combining in quadrature the errors on the branching and fit fractions in equation \rf{eq:ff}.

A striking feature of the table is the magnitude of the fit fractions in the first two columns compared to all the others. This is mainly because the corresponding three-body branching fractions $\brf(B\to D\D K)$ are small compared to the others, which can in turn be partly understood by the flavour topology. Whereas the three-body transitions $B^+\to\ddka$ and $B^0\to\ddkd$ cannot be produced directly via the colour-favoured mechanism, the other four transitions can. This is apparent from equations \rf{eq:phip} and \rf{eq:phin}, and is also discussed in Ref.~\cite{delAmoSanchez:2010pg}. 

The dramatic prominence of the  experimental $X(2900)$ peak may be understood in this context. We have shown that  the $B^+\to D^+X, X\to \dka$ transition can occur through colour-favoured processes, either through triangle diagrams or resonant $X$ states. But the experimental background is comparatively small, because the direct process $B^+\to\ddka$ is colour-suppressed. The combination of production through colour-favoured processes with a background which is colour-suppressed implies a large fit fraction, hence a prominent experimental signal.

With this observation in mind, a comment on our starting assumption is in order. In setting up our model we have ignored the contribution from colour-suppressed decays, on the basis that in two-body $B$ decays they are {much smaller} (by around a factor of 10) in comparison to colour-favoured decays. Naively, the numbers in  Table~\ref{tab:fit} may suggest that the suppression effect is less substantial in three-body decays, but this is misleading. As noted, three-body final states whose direct production is colour-suppressed can nonetheless arise indirectly through colour-favoured processes, either through intermediate resonances or triangle diagrams, and these indirect processes can account for a substantial part of the three-body branching fraction; as an example, in the LHCb amplitude analysis~\cite{Aaij:2020ypa}, the $X(2900)$ states account for a larger fit fraction than the non-resonant $\ddka$ component. Two-body $B$ decays are, by comparison, much easier to interpret, and the clear evidence for substantial colour suppression in those cases justifies our original assumption.

An obvious message of Table~\ref{tab:fit} is that the best channel for studying neutral $\D K$ resonances is very likely to be the one already analysed by LHCb. The experimental fractions for the discovery mode (first column) are larger than the predicted fractions for other  modes for the neutral $X(2900)$ states, except in the case that a particular choice of parameters ($C_0$, $C_1$ or $\theta$) render modes expressed in terms of these to be larger.

However we also predict, regardless of the model, a very prominent experimental signal in the $\bdxd$ mode (second column). The prediction follows from the general result \rf{eq:gen}, which relies only the assumption that production is dominated by colour-favoured transitions.  Since all of our other predictions rely on the same assumption, experimental study of  $\bdxd$ would be a critical test of model assumptions.

The predictions in the remaining columns can be used to discriminate among models. Although the numbers are generally smaller, in considering the experimental feasibility of observing these modes, we note that it is quite standard  in amplitude analyses to resolve structures with a fit fraction at level of a few percent -- the $X_0(2900)$ is one such example. 

Our discussion concentrates on models with an explicit prediction for the fit fraction, as opposed to those expressed in terms of unknown parameters $C_0$, $C_1$ or $\theta$. Our expressions for the latter cases may be used to constrain parameters by comparison with future experimental data. {We comment below on other possible experimental and theoretical constraints on these parameters.}

For the other neutral $X(2900)$ modes (columns 3 and~4) we first recall the striking prediction that in the triangle scenario with quark exchange, these modes would be absent. For the other scenarios, even considering the significant uncertainties ($35\%$) on the numbers, the predicted fit fractions in the triangle (OPE) and resonance ($I=0$ or $I=1$) interpretations are not consistent with one another. Hence experiment may be able to discriminate between these possibilities. (The factor of 4 relating the resonance and triangle scenarios is due to $R$, see Table~\ref{tab:r}.)

The previous observations can be useful even if only one or the other of the modes in columns 3 or 4 is measured in experiment. Measurement of both modes would be even more revealing. Apart from the case of a resonance with mixed isospin, the two modes satisfy equation~\rf{eq:web}, and so have the same $R$. The relation between the modes is then a fixed numerical factor (= 1.35) which comes from the other terms in equation~\rf{eq:ff}. Measured fit fractions which are consistent with this ratio, but not with the predictions for the  triangle (OPE) or resonance ($I=0$ or $I=1)$ scenarios, support the triangle~(EFT) scenario.  On the other hand, fit fractions which are not consistent with this ratio would indicate a resonance of mixed isospin. In principle, the mixing angle $\theta$ could be extracted from the measured ratio.

We now move to the final two columns, for the charged partners of the $X(2900)$ states. Absence of these modes would be a striking signature of the $I=0$ resonance scenario. In all other scenarios the charged modes are expected, and since their predicted magnitudes differ considerably between models, experimental measurement of their fit fractions can discriminate among models.  (The predicted numbers do {not} overlap despite the significant uncertainties.) Particularly noteworthy are the very large fit fractions  in the $I=1$ resonance scenario. If the $X(2900)$ states are isovector resonances, as advocated in some models, they would be extremely prominent in $\bdxc$ and $\bdxf$.

{We finally remark on modes which are sensitive to the parameters $C_0$, $C_1$ or $\theta$. As noted previously, future experimental measurement of these modes can give constraints on the parameters, using the expressions in Table~\ref{tab:fit}. Alternatively, if the parameters can be obtained from elsewhere, the expressions in Table~\ref{tab:fit} can be used to give predictions for experiment. With this in mind, we now give some suggestions for how the ratio $C_0/C_1$, and the mixing angle $\theta$, may be obtained from other experiments, or from theory. }

{In principle, the ratio $C_0/C_1$ could be determined from lattice QCD studies of relevant scattering process in several charge channels, such as $\D^{*0}K^{*0}\to D^- K^+$ and $\D^{*0}K^{*0}\to \D^0 K^0$. These inelastic scattering processes have not previously been studied in lattice QCD; elastic $\bar D K$ scattering is discussed in refs.~\cite{Liu:2012zya,Cheung:2020mql}. The ratio $C_0/C_1$ could also perhaps be deduced from the quark model, where meson scattering is described in terms of the exchange of light mesons (such as $\pi$, $\rho$, $\eta$).
We emphasise that predictions involving $C_0/C_1$ only apply if the interactions are perturbative.}

{The mixing angle $\theta$, which applies in the resonance scenario, could be determined experimentally from the relative branching fractions of $X(2900)$ decays in various charge modes. Owing to the factorisation discussed previously, in the resonance scenario the relative decay branching fractions are independent of the production mechanism, hence this approach could be useful not only in $B$ decays. The ratio of
$\dka$ and $\dkb$ modes fixes the mixing angle according to equation~\rf{eq:rc}, and similarly for the corresponding charge combinations in $\D^*K$ and $\D K^*$. The angle could also be extracted from the relative fractions of different charge combinations in three-body final states $\D K\pi$ and $\D^*K\pi$. It may also be possible to predict $\theta$ in models, for example in the molecular scenario where mixing arises from isospin-violating mass differences among the constituents; the approach is however limited by the uncertainties on the $X(2900)$ and $K^{*+}$ masses.}

\section{Two-body branching fractions}
\label{sec:tb}

Finally we give some additional results which apply only to the resonance scenario. In this case, as noted in Section~\ref{sec:trd}, the matrix element for $B\to DX, X\to \D K$ factorises into a product of matrix elements for production ($B\to D X$) and decay ($X\to \D K$), which implies a corresponding factorisation of the branching fractions, equation \rf{eq:factorise}.

We emphasise that this factorisation does not apply in the triangle scenario. It makes no sense physically to separate ``production'' ($\D^* K$ or $\D_1K$ loops) and ``decay'' (the $D^-K^+$ final state), considering that the origin of the signal is exactly the interplay of these two processes. We note also that attempting to impose such a factorisation on the matrix elements leads immediately to algebraic problems. An extreme example is the triangle scenario with QE, in which the zero entry in the middle column of Table~\ref{tab:me} implies that one or the other of the matrix elements for  $B^+\to D^+ X$ or $X\to\dkb$ is zero, and similarly one or the other of $B^0\to D^0 X$ or $X\to\dka$ is zero. This is obviously inconsistent with the non-zero entry in the left column. More generally, by considering the general parametrisation of matrix elements from the top row of Table~\ref{tab:me}, it is easy to see that with the assumption of factorisation, the expressions are only self-consistent if $\lambda_0=0$ or $\lambda_1=0$, corresponding to the $I=1$ and $I=0$ resonance scenarios. The mixed isospin states are also obviously consistent with factorisation. The conclusion is that only resonance scenarios are consistent with a factorisation of matrix elements, hence branching fractions.

\begin{figure*}
\includegraphics[width=\textwidth]{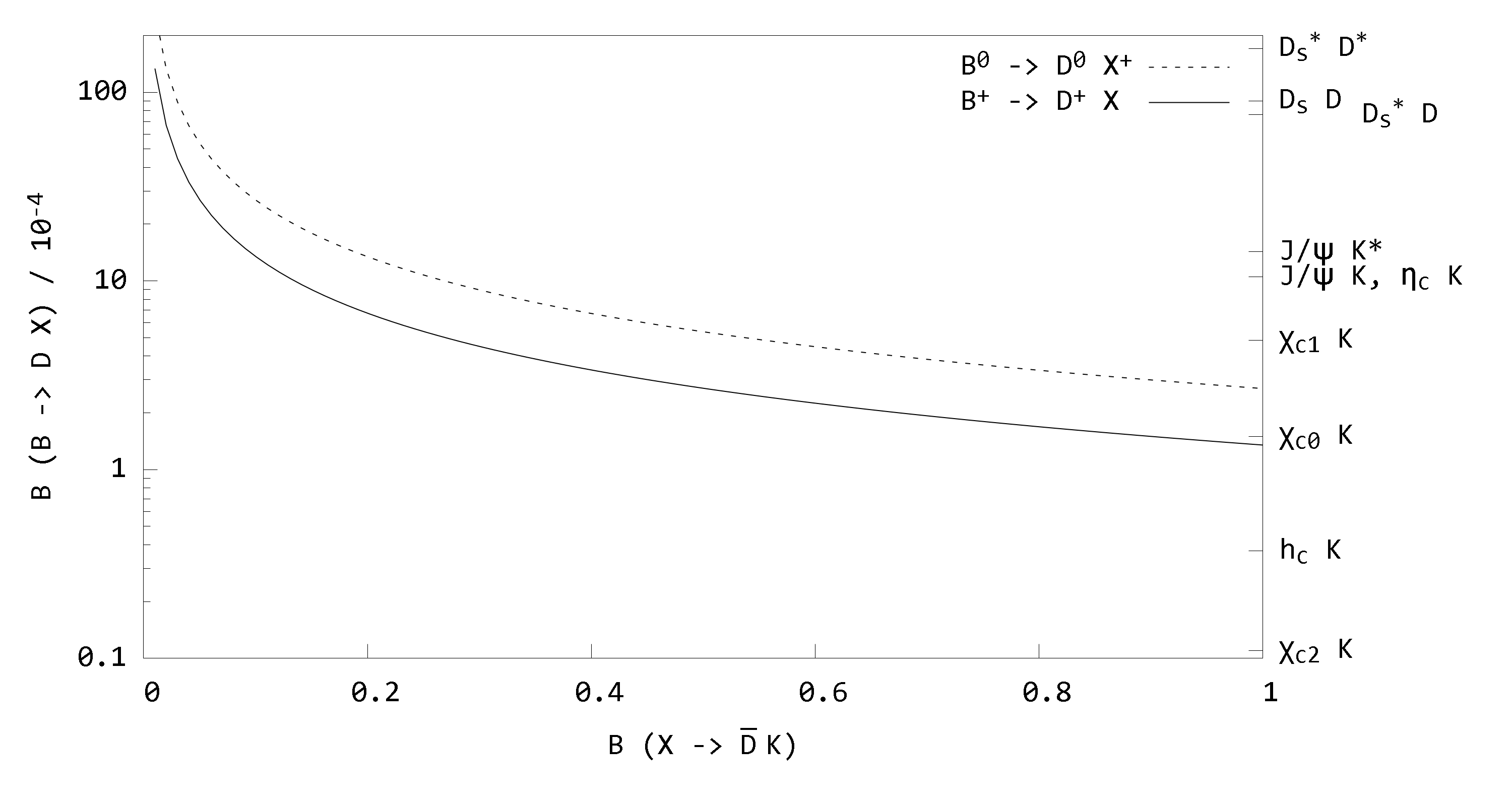}
\caption{Production branching fractions for the $X_1(2900)$ state and its possible charged partner, in the $I=0$ or $I=1$ resonance scenarios. The vertical axis shows   $\brf(B^+\to D^+ X)$ (solid) and $\brf(B^+\to D^0 X^+)$ (dashed), in units of $10^{-4}$. The horizontal axis is the total  $\D K$ decay branching fraction $\brf(X\to \D K)$ of the neutral $X_1(2900)$.}
\label{fig:bf}
\end{figure*}

We thus concentrate exclusively on the resonance scenarios in this section. Since the branching fractions factorise, in addition to the previous relations among $\brf(B\to DX, X\to \D K)$, we get further relations among $\brf(B\to DX)$ and $\brf(X\to \D K)$ separately. These follow trivially from equations~\rf{eq:nc}-\rf{eq:rc}.

Using these, we can get lower limits on $\brf(B\to DX)$. For concreteness, we assume either $I=0$ or $I=1$; it is easy to generalise to the case of mixed isospin which is, however, less predictive.

For the neutral $X(2900)$ states, the total $\D K$ branching fraction is
\begin{multline}
\brf(X\to \D K)\\=\brf(X\to \dka)+\brf(X\to \dkb).
\end{multline}
 Ignoring small differences due to phase space, the fractions on the right-hand side are equal, implying
\begin{align}
\brf(X\to \dka)=\frac{1}{2}\brf(X\to \D K).
\end{align}
Using this in equation~\rf{eq:bft}, having factorised the branching fractions on the left-hand side, we get the production branching fraction for $X_1(2900)$, 
\begin{align}
\brf(B^+\to D^+ X)=\frac{(1.35\pm 0.32\pm 0.32)\times 10^{-4}}{\brf(X\to \D K)}.\label{eq:bfa}
\end{align}
Note that, unlike most of our other results, this does not rely on the assumed dominance of colour-favoured processes. There is a similar prediction for $\brf(B^0\to D^0 X)$, using \rf{eq:nc} and adjusting for the $B^0$ lifetime. Results for $X_0(2900)$ are smaller by 5.6/30.6. 

If the neutral $X_1(2900)$ belongs to an $I=1$ multiplet, from equation~\rf{eq:ncb}, the production branching fraction for its charged counterpart is
\begin{align}
\brf(B^+\to D^0 X^+)=\frac{(2.69\pm 0.53\pm 0.52)\times 10^{-4}}{\brf(X\to \D K)},\label{eq:bfb}
\end{align}
where we emphasise that the decay branching fraction in the denominator is that of the neutral $X_1(2900)$. Again, there are similar results for $B^0$ decays, and for $X_0(2900)$.

The preceding equations imply lower limits
\begin{align}
\brf(B^+\to D^+ X)&>{(1.35\pm 0.32\pm 0.32)\times 10^{-4}},\\
\brf(B^+\to D^0 X^+)&>{(2.69\pm 0.53\pm 0.52)\times 10^{-4}},
\end{align}
meaning that the production branching fractions are at least a factor of two larger than those quoted in  Ref.~\cite{Chen:2020eyu}.

In Figure~\ref{fig:bf} we plot equations \rf{eq:bfa} and \rf{eq:bfb}, showing
$\brf(B^+\to D^+ X)$ (solid line) and $\brf(B^+\to D^0 X^+)$ (dashed) as a function of the total  $\D K$ decay branching fraction $\brf(X\to \D K)$ of the neutral $X_1(2900)$. On the right-hand side of the plot we show some examples of  production branching fractions for conventional mesons which are also produced via Cabibbo-favoured transitions. It is remarkable that, in the resonance scenario,  the exotic $X(2900)$ states and their possible charged partners are produced at least as copiously as many conventional mesons.


With reference to the plot, note it would be quite natural to have $\brf(X\to \D K)\ll 1$, implying very large production branching fractions $\brf(B\to D X)$. This is because the quantum numbers of $X_1(2900)$ allow for several decays other than $\D K$. In particular, there are two-body modes $\D^* K$, $\D K^*$, and several three-body modes including $\D K\pi$ and $\D^* K\pi$. (The last of these could arise from the decay of the $K^*$ constituent \cite{Huang:2020ptc}.) We also note that the observed $\D K$ mode is P-wave, whereas the three-body modes are S-wave, suggesting they could account for significant branching fraction.

In the model of Ref.~\cite{Huang:2020ptc}, the $\D K$ partial widths of $X_1(2900)$ are tiny. They find it impossible to reconcile these with the $X_1(2900)$ width and use this as an argument against the molecular interpretation. Even if the missing width could be explained in some other way, there would be a further problem. The $\brf(X\to \D K)$ fractions are ${\mathcal O} (10^{-5})$, which is clearly impossible considering equation \rf{eq:bfa}. 

Another intriguing comparison is to $X(3872)$:
\begin{align}
\brf(B\to X(3872)K^+)<2.6\times 10^{-4}.
\end{align} Amongst exotic hadrons, $X(3872)$ is the one which has been studied most thoroughly in experiment. Yet the $X_1(2900)$ has at least comparable, very likely larger, production branching fraction, suggesting rich possibilities for further experimental study.

Production branching fractions for $X_0(2900)$ will be somewhat smaller. This is because, as well as the suppression by a factor 5.6/36, we expect $\brf(X\to \D K)$ to be larger than in the $X_1(2900)$ case. The rationale is that the observed $\D K$ mode is the only two-body decay allowed by spin-parity, and since it is S-wave, we expect it to account for significant branching fraction. 
Additionally, since a scalar cannot decay to three pseudoscalars, the $\D K\pi$ mode which is possible for $X_1(2900)$ is not possible for $X_0(2900)$. The model calculation of Ref.~\cite{Huang:2020ptc} finds $\brf(X\to \D K)\approx 0.9$. Even with this large number, the resulting production branching fractions (see Fig.~\ref{fig:bf}) are still in excess of several conventional mesons.

\section{Conclusions}
\label{sec:conclusion}

Understanding the nature of the $X(2900)$ states requires further experimental study in other production and decay modes, and a search for their charged partners. 

Intriguingly, we find that for the neutral $X(2900)$ states, the channel with the largest fit fraction is the discovery mode $\bdxa$. This is because the $X(2900)$ signal can arise through a colour-favoured process, mediated by triangle diagrams or resonances, whereas the experimental background is colour-suppressed.

Another mode which is large for the same reason is $\bdxd$. We predict a significant fit fraction of around 23\%, regardless of the nature of the $X(2900)$ states. Confronting this prediction with experiment would be a useful test of the central idea in our approach, which is that production is dominated by colour-favoured processes.

Among the remaining modes, we have shown that there are characteristic patterns in production and decay which discriminate unambiguously between competing models. 

The triangle scenario with quark exchange is characterised by the striking prediction that in $B^+\to D^+ X$, the $X(2900)$ states are seen in $\dka$ but not $\dkb$, whereas in $B^0\to D^0 X$ the pattern reverses, with the states seen in $\dkb$ but not in $\dka$. The modes which are forbidden in this scenario are allowed in the alternative triangle scenario where the interactions are based on OPE. Their fit fractions, however, would be smaller than in the resonance scenario. The two triangle scenarios both imply a charged partner state, but the predicted fit fractions are  sufficiently different from each other, and from the resonance scenario, that experiment could discriminate among the models.

In the resonance scenario, the neutral $X(2900)$ states have the same fit fractions regardless of whether they are $I=0$ or $I=1$. The two possibilities would instead be distinguished by the existence of a charged partner in the latter case, which has  enormous fit fractions even exceeding that of the observed $X_1(2900)$ in its discovery mode. In this context we suggest  the experimental study of $\bdxc$ and $\bdxf$. If the $X(2900)$ states belong to an isotriplet, their charged partners would be abundant in these modes.

The neutral $X(2900)$ states could alternatively have mixed isospin, although this will be difficult to establish experimentally, unless the mixing angle is very large. The modes $\bdxd$ and $\bdxe$ are useful in this context; deviation of the ratio of their fit fractions from 1.35 would be a ``smoking gun'' of a resonance with mixed isospin.

Finally we note that in the resonance scenario, where production and decay factorise, the production branching fractions of $\brf(X\to \D K)$ are very large, comparable to those of conventional mesons and larger than that of another exotic state, $X(3872)$.

\acknowledgments

Swanson's research was supported by the U.S. Department of Energy under contract DE-SC0019232.

\bibliography{C:/Users/bernh/Documents/LaTeX/bibinputs/tjb}
\end{document}